\DocumentMetadata{}
\documentclass[sigconf]{acmart}

\makeatletter
\def\@ACM@checkaffil{% Only warningsd  
    \if@ACM@instpresent\else
    \ClassWarningNoLine{\@classname}{No institution present for an affiliation}%
    \fi
    \if@ACM@countrypresent\else
        \ClassWarningNoLine{\@classname}{No country present for an affiliation}%
    \fi
}
\makeatother

\AtBeginDocument{%
  \providecommand\BibTeX{{%
    \normalfont B\kern-0.5em{\scshape i\kern-0.25em b}\kern-0.8em\TeX}}}

\usepackage{xspace}
\usepackage{balance}
\usepackage{romannum}
\usepackage{color, soul, xcolor}

\def\etal{\textit{et~al.}\xspace}

\def\eg{\textit{e.g.,}\xspace}

\copyrightyear{2025}
\acmYear{2025}
\setcopyright{rightsretained}
\acmConference[UIST Adjunct '25]{The 38th Annual ACM Symposium on User Interface Software and Technology}{September 28-October 1, 2025}{Busan, Republic of Korea}
\acmBooktitle{The 38th Annual ACM Symposium on User Interface Software and Technology (UIST Adjunct '25), September 28-October 1, 2025, Busan, Republic of Korea}\acmDOI{10.1145/3746058.3758372}
\acmISBN{979-8-4007-2036-9/2025/09}

\author{Nishanth Chidambaram}
\orcid{0009-0008-2729-5626}
\affiliation{%
  \department{Computer Science and Engineering}
  \institution{University of California San Diego}
}
\email{nchidambaram@ucsd.edu}

\author{Weichen Liu}
\orcid{0000-0002-8576-6130}
\affiliation{%
  \department{Computer Science and Engineering}
  \institution{University of California San Diego}
}
\email{wel008@ucsd.edu}
\authornote{Both authors contributed equally to the paper}

\author{Manas Satish Bedmutha}
\orcid{0000-0003-3427-2226}
\affiliation{%
  \department{Computer Science and Engineering}
  \institution{University of California San Diego}
}
\email{mbedmutha@ucsd.edu}
\authornotemark[1]

\author{Nadir Weibel}
\orcid{0000-0002-3457-4227}
\affiliation{%
  \department{Computer Science and Engineering}
  \institution{University of California San Diego}
}
\email{weibel@ucsd.edu}

\author{Chen Chen}
\orcid{0000-0001-7179-0861}
\affiliation{%
  \department{Computing and Information Sciences}
  \institution{Florida International University}
}
\email{chechen@fiu.edu}

\begin{document}

\def\sysname{DriveSimQuest}

\title{\sysname: A VR Driving Simulator and Research Platform on Meta Quest with Unity}

\begin{abstract}
Using head-mounted Virtual Reality (VR) displays to simulate driving is critical to study driving behavior and design driver assistance systems. But existing VR driving simulators are often limited to tracking only eye movements. The bulky outside-in tracking setup and Unreal-based architecture also present significant engineering challenges for interaction researchers and practitioners. We present {\it \sysname}, a VR driving simulator and research platform built on the Meta Quest Pro and Unity, capable of capturing rich behavioral signals such as gaze, facial expressions, hand activities, and full-body gestures in real-time. \sysname~ offers a preliminary, easy-to-deploy platform that supports researchers and practitioners in studying drivers’ affective states and behaviors, and in designing future context-aware driving assistance systems.
\end{abstract}

\maketitle

\section{Introduction}

\begin{figure}
    \centering
    \includegraphics[width=.99\columnwidth]{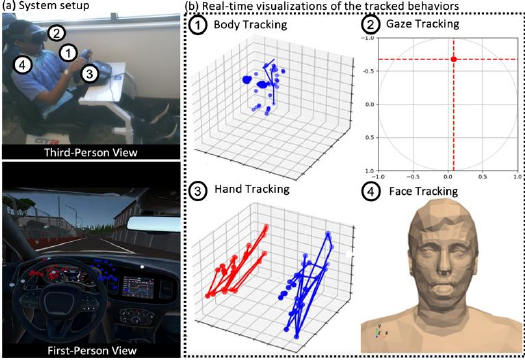}
    \vspace{-0.15in}
    \caption{Overview of \sysname; (a) system setup; (b) example real-time visualizations of behavior tracking, \textit{incl.} body (b1), gaze (b2), hand (b3), and reconstructed face (b4).}\
    \vspace{-0.1in}
    \label{fig::system}
\end{figure}

Using head-mounted Virtual Reality (VR) displays to simulate driving experiences is a widely adopted approach for designing driving assistance systems, as the virtual environment can closely mirror participants' experiences in real-world settings~\cite{Slater2006, Lee1998} with added values ~\cite{Pettersson2019}. Additionally, driving in VR enables more controlled and safer experiments, with less cumbersome setups for data collection~\cite{Winter2012}. A VR driving simulator also facilitates interaction research and simulations on emerging visual display systems - such as in-car displays and heads-up displays \cite{Maroto2018} - which are indispensable components of advanced driver assistance systems.

While prior works have explored VR driving experiences, the simulators used in these works often rely on outside-in tracking sensors such as HTC Vive's lighthouse \cite{Vive} and Oculus Rift's constellation cameras \cite{OculusRiftHistory} - which demand additional bulky hardware setup in the experimental space. In addition, existing VR driving simulators often lack the capability to capture rich eye metrics, facial expression, and full bodily gestures from driver participants, which are often considered indispensable cues for a broader behavioral and interaction research~\cite{Zoller2018, Agarkar2023, Aftab2019, Chen2021}. For example, Pettersson~\etal~\cite{Pettersson2019} compared the VR driving experience with real-world driving using a simulator; while they found VR promising for evaluating in-vehicle systems, they also noted limitations in realism and a need to integrate physical sensations to increase the ecological validity of VR-based studies. Lhemedu-Steinke~\etal~\cite{IhemeduSteinke2017, LhemeduSteinke2018} built a VR driving simulation system using Oculus Rift \cite{OculusRiftHistory}, but they were not able to perform any affective and embodied driver analysis key to understand behavior. DReyeVR~\cite{Silvera2022DReyeVR} demonstrated an Unreal-based VR driving simulation system that can be rendered on Meta Quest 2~\cite{Quest2} and HTC Vive~\cite{Vive} for behavioral and interaction research based on Carla~\cite{Dosovitskiy2017Carla}; despite the support for hand tracking and eye tracking via the HTC Vive Pro Eye \cite{ViveProEye}, it remained challenging to capture facial expressions and full-body gestures. Finally, most of the existing systems are built on Unreal Engine, which can introduce unnecessary prototyping challenges for researchers~\cite{Soni2024, berrezueta2025immersive}.

In this work, we present \textit{\sysname} (Figure~\ref{fig::system}), a VR driving simulator and research platform built upon the Logitech G923 racing wheel and pedals~\cite{g923} (Figure \ref{fig::system}a) and developed using Meta Quest Pro~\cite{MetaQuestPro} and Unity, capable of tracking key bodily behaviors, including gaze, facial expressions, and body movements. Figure \ref{fig::system}b shows an example visualization of the tracking results. Unlike prior works (\eg~\cite{Silvera2022DReyeVR, IhemeduSteinke2017, LhemeduSteinke2018, Pettersson2019}), \sysname~ offers a preliminary, easy-to-deploy platform that supports researchers and practitioners in studying drivers' affective states and behaviors, and in designing future context-aware driving assistance systems. We believe that, \mbox{\sysname} can serve as a research platform for future work on designing interactive, empathic and driver-passenger(s)-aware driving assistant interfaces.

\section{System Design}

Fig.~\ref{fig::system}a shows an overview of \sysname's setup. We describe its key design components below.

\vspace{2px}\noindent{\bf Hardware.}
\sysname~ uses the Quest Pro~\cite{MetaQuestPro} as the VR headset, due to its inside-out tracking and rich full-body tracking capabilities. While the current \sysname~ prototype uses the Quest Pro~\cite{MetaQuestPro}, future \sysname~ can be easily upgraded to newer headsets that offer similar gaze, facial, and body tracking capabilities. \sysname~ also integrates the Logitech G923 racing wheel and pedals~\cite{g923} to provide tangible and realistic driving input.

\vspace{2px}\noindent{\bf Behavior tracking.}
\sysname~ leverages Meta Quest Pro's built-in inside-out camera and IMU setup to enable gaze, facial, and full body tracking, made possible through integration with the Meta Movement SDK~\cite{MovementSDK}.
In the future, researchers will be able to efficiently access the transform data of key hand and body joints, alongside eye gaze rays and $70$ different facial blendshape weights.
While most prior research has focused primarily on gaze \cite{Silvera2022DReyeVR}, \mbox{\sysname} expands these capabilities by enabling researchers to explore and leverage implicit behaviors expressed through facial expressions, hand and full-body gestures.

\vspace{2px}\noindent{\bf A Unity-based architecture.}
While most existing driving simulation engines (\eg~\cite{Silvera2022DReyeVR}) are built with Unreal, we chose Unity for its flexibility and lower learning curve~\cite{Soni2024}, making it more accessible for interaction and behavior researchers as well as practitioners.
The engine behind \sysname ~ is repurposed from AirSim~\cite{Shah2017AirSim} - an open-sourced high-fidelity visual and physical simulation engine for driving and drone applications.
While AirSim is designed for traditional display-based applications, \sysname~ extends its capabilities to Quest Pro, by integrating Meta All-In-One SDK~\cite{MetaAllInOneSDK}.
Although a PC VR setup could better support high-end graphics, we prioritized a standalone design to maximize access to the Meta Quest Pro’s built-in behavior tracking sensors and simplify deployment.
\sysname~ then serializes and broadcasts the tracking data and input actions (\eg~ steering and throttle level) through UDP, which can be easily accessed, analyzed and integrated by researchers and developers as part of external applications.

\vspace{2px}\noindent{\bf Real-time visualizations.}
For demonstration purposes, the current deployment of \sysname~ includes an external application that continuously retrieve and visualizes the behavior tracking data and time-series input actions in real-time (Fig.~\ref{fig::system}b).

\section{Example Analysis}
We present two example analysis drawn from our preliminary exploratory studies, where participants drove with \sysname. While our ultimate goal is to design a context-aware driving assistant experience, this work seeks to gain initial insights into how tracked data can be leveraged to understand driver behavior in specific contexts. Comprehensive human-factor evaluations of the tool usability are planned as part of future work.

\begin{figure}[t]
    \vspace{-0.1in}
    \centering
    \includegraphics[width=\columnwidth]{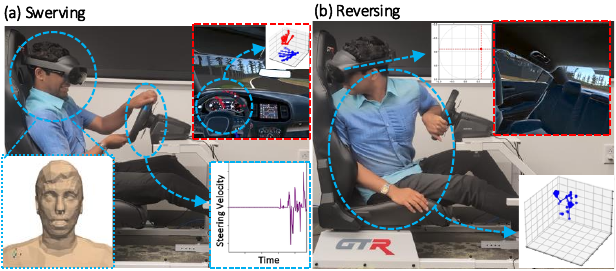}
    \vspace{-0.25in}
    \caption{Exploratory Analysis of two example scenarios; (a) visualizations of facial expressions, hand movements, and steering velocity during swerving; (b) visualizations of captured gaze and body pose during reversing. First-person views are shown at the top right of (a - b).}
    \vspace{-0.1in}
    \label{fig::eval}
\end{figure}

The first example focuses on the driver’s facial expressions and hand poses during swerving in an emergent turning scenarios. Driver's fear can be visually inferred from tracked facial expressions, referenced against the driver’s real face while wearing the VR headset (Figure~\ref{fig::eval}a). We also visualize the captured steering velocity (Figure~\ref{fig::eval}a bottom right) to support interpretation of the driver's actions in response to such an emergent scenario.

Our second example highlights a driver's behavior while reversing the car -- a task that often involves coordinated gaze and body movements, as the driver may need to physically turn around to check blind spots for clearance, for instance in the context of parallel parking. Figure \ref{fig::eval}b, illustrates how upper body gestures are tracked and visualized in reference to the actual body movements.

\section{Conclusion and Future Work}
\sysname~is a VR driving simulator and research platform based on Meta Quest Pro and developed in Unity that allows researchers to access, understand and integrate tracked behavior data like gaze, face, hand and body.
\sysname~ paves a new way for interaction researchers to easily gain deeper insights into current driving experiences and to design more effective ones in the future. 
Alongside our continuous development and evaluations of \sysname ~(\eg~ a thorough user study aimed at understanding developers’ experiences while working with \sysname), the directions of our long-term future work is twofold.

\vspace{2px}\noindent\textbf{Driving behavior analysis.} The first direction involves leveraging \sysname's behavior tracking data to gain deeper insights into drivers' behavior.
\sysname~goes beyond understanding behavior through eye movements~\cite{Silvera2022DReyeVR}, and will allow researchers to gain deeper insights into facial expressions, hands and full-body posture.
This will for example enable analysis of drivers' confusion -- often detectable by tracking facial gestures  -- and possibly inform the design of road signs and how could driving assistance systems, such as car play, better present information.
\sysname~ also enables analysis of specific driving behaviors, for example by assessing driving performance during parallel parking using tracked body and hand movements.

\vspace{2px}\noindent\textbf{Designing context-aware driving assistant systems.}
A second line of future work could focus on designing context-aware driving assistance experiences. 
While existing research~\cite{Silvera2022DReyeVR} primarily focuses on driver attention (which is often inferred through attention and gaze), \sysname~ enables incorporating richer contextual cues, such as affective states expressed through facial expressions and body gestures.
For example, this would allow \textit{in-situ} level of detail on the car display to be computationally adapted based on tracked behavioral data during complex driving scenarios involving route decisions; facial expressions could reveal signs of confusion, while hand and gaze patterns could help infer the specific spatial elements causing uncertainty.

\begin{acks}
We acknowledge the feedback provided during the early-stage idea brainstorming from Aaron Broukhim, as well as suggestions from the anonymous reviewers.
\end{acks}

\balance
\bibliographystyle{ACM-Reference-Format}
\bibliography{references}

\end{document}